\PassOptionsToPackage{unicode=true}{hyperref} 
\PassOptionsToPackage{hyphens}{url}
\documentclass[]{article}
\usepackage{graphicx}

\usepackage{lmodern}
\usepackage{amssymb,amsmath}
\usepackage{ifxetex,ifluatex}
\usepackage{fixltx2e} 
\ifnum 0\ifxetex 1\fi\ifluatex 1\fi=0 
  \usepackage[T1]{fontenc}
  \usepackage[utf8]{inputenc}
  \usepackage{textcomp} 
\else 
  \usepackage{unicode-math}
  \defaultfontfeatures{Ligatures=TeX,Scale=MatchLowercase}
\fi
\IfFileExists{upquote.sty}{\usepackage{upquote}}{}
\IfFileExists{microtype.sty}{%
\usepackage[]{microtype}
\UseMicrotypeSet[protrusion]{basicmath} 
}{}
\IfFileExists{parskip.sty}{%
\usepackage{parskip}
}{
\setlength{\parindent}{0pt}
\setlength{\parskip}{6pt plus 2pt minus 1pt}
}
\usepackage{hyperref}
\hypersetup{
            pdfborder={0 0 0},
            breaklinks=true}
\usepackage{longtable,booktabs}
\IfFileExists{footnote.sty}{\usepackage{footnote}\makesavenoteenv{longtable}}{}
\ifx\paragraph\undefined\else
\let\oldparagraph\paragraph
\renewcommand{\paragraph}[1]{\oldparagraph{#1}\mbox{}}
\fi
\ifx\subparagraph\undefined\else
\let\oldsubparagraph\subparagraph
\renewcommand{\subparagraph}[1]{\oldsubparagraph{#1}\mbox{}}
\fi

\makeatletter
\def\fps@figure{htbp}
\makeatother

\date{}

 \usepackage{fancyhdr}

\begin{document}

\textbf{Trait-based indices to assess benthic vulnerability to trawling and model potential loss of ecosystem functions} \footnote{© 2020. This manuscript version is made available under the CC-BY-NC-ND 4.0 license \href{http://creativecommons.org/licenses/by-nc-nd/4.0/}{http://creativecommons.org/licenses/by-nc-nd/4.0/}}

Hilmar Hinz\textsuperscript{a}, Anna Törnroos\textsuperscript{b} and
Silvia de Juan\textsuperscript{c}

\textsuperscript{}\\

\textsuperscript{a}Mediterranean Institute for Advanced Studies
(IMEDEA), Department of Ecology and Marine Resources Esporles, Balearic
Islands, Spain

\textsuperscript{b}Åbo Akademi University, Environmental and Marine
Biology, Department of Biosciences, Turku, Finland

\textsuperscript{c}Institut de Ciències del Mar (ICM), Renewable Marine
Resources, Barcelona, Spain

\textbf{Corresponding Author:}

Hilmar Hinz; email: hhinz@imedea.uib-es


\textbf{Abstract}

The physical impact of bottom towed fishing gears does not only reduce
the abundance and biomass of species, but also alter the overall
composition and through this functioning of benthic communities. The
vulnerability to trawling on a species level is determined by a species'
individual combination of morphological, behavioural and life history
traits. In turn, the ecosystem functions most affected will be defined
by those same species, and their contribution towards a particular
function. On the basis of this paradigm, trait-based indices of physical
resistance (\textbf{RI}) and reproductive potential (\textbf{RPI}) were
developed and combined into an overall vulnerability index on a species
level, the \textbf{RRI} or \textbf{R}esistance and \textbf{R}eproductive
Potential \textbf{I}ndex. The developed indices can be used to explore
how resistance and reproductive potential, as well as overall community
vulnerability, change over a trawling gradient. Furthermore, the RRI
allows for dividing the benthic community into groups expressing
different levels of vulnerability that can be linked to ecosystem
functions to explore functional vulnerability to trawling. The
\textbf{RRI} index furthermore opens up the possibility of scenario
modelling by simulating the extinction or loss of vulnerable species and
its effects on functions. This may be of particular interest in data
poor case studies where there is lack of data over trawling gradients,
or to project and explore the consequences of increases in fishing
effort in a particular location. The validity of the trait-based
\textbf{RRI} index was explored by comparing individual species' RRI
scores to empirically observed responses over a trawling gradient based
on a previously published data. \textbf{RRI} score and observed
responses (regression slopes) were significantly correlated providing
support for the rationality of the approach taken. Moreover, further
analysis of the data evidenced clear increases of resistance and
resilience indices over the trawling gradient, demonstrating that
communities lost vulnerable species with increasing trawling. When
exploring the effects of trawling on the bioturbation as a chosen
ecosystem function, we found it to be disproportionately affected though
the loss of vulnerable species.{ }The proposed indices provide new
insights into the link of species vulnerability and function. Such
information is of vital interest to environmental managers focused on
persevering ecosystem functions and services in the face of
anthropogenic global change.

\textbf{1. Introduction}

Globally, fishing impacts represent one of the main anthropogenic
pressures acting on the marine environment (Clark et al., 2016; Eigaard
et al., 2017) with negative consequences for the productivity and
functioning of locally affected benthic ecosystems (Hiddink et al.,
2011; Olsgard et al., 2008; Queirós et al., 2006). To safeguard the
integrity of benthic ecosystems from fishing related impacts, various
policy initiatives (e.g. among others Magnuson Stevenson Act USA,
European Marine Strategy Framework Directive EU) have been promoted
within the context of an ecosystem-based approach to fisheries
management (Berg et al., 2015; Biedron and Knuth, 2016; Garcia et al.,
2003). However, to evaluate the effectiveness of regulatory
interventions, it is essential to monitor the health or status of
benthic habitats through indicators that are able to capture changes in
the structure and functioning of benthic ecosystems (de Juan et al.,
2015; Rice et al., 2012).

Indicators that measure the status of benthic habitats have primarily
been based on community metrics such as density, biomass and diversity
(Hiddink et al., 2020) of the entire or of a specific size fraction of
the benthos (McLaverty et al., 2020). A recent meta-analysis by Hiddink
et al. (2020) concluded that community biomass as an indicator of
fishing impacts gave the most reliable consistent responses compared to
other community indicators. The disadvantage of these generalized
community indicators is that, on their own, they cannot provide
estimates of the functional status or loss associated with fishing
impacts, as biomass and functioning may be reduced at different rates
(Thrush et al., 2006). Thus, which functions are most affected by
trawling and to which degree, will be defined by the biomass composition
of species, their contribution towards a function and, most importantly,
their vulnerability to trawling. The vulnerability of a species will in
turn depend on the morphology, behaviour and life history
characteristics of the species (Bremner et al., 2006; de Juan et al.,
2020; Juan et al., 2007). Therefore, an ecosystem function will be most
affected by human activities if the species that contribute
significantly to that function are also the most vulnerable to the
respective activity. Henceforth, to understand and potentially predict
the consequences of external drivers of ecosystem change from trawling
on benthic functioning, it is essential to integrate ecological
information at a species level.{~}

With this realisation in mind, benthic ecologists have implemented the
concept of biological traits to assess the wider ecosystem consequences
of changes in marine community composition (Bremner et al., 2003;
Törnroos et al., 2019). Hereby, species' biological attributes that
describe certain aspects of their morphology and behaviour are used with
the aim to approximate the ecological role of the species (Bremner et
al., 2003; Törnroos and Bonsdorff, 2012). Simple examples are
categorisations into morphological attributes such as size or fragility
(Shin et al., 2005), while others are related to behaviour, for example
mobility and feeding mode (Smale, 2008), and to life history traits,
such as maximum size, fecundity or similar (King and McFarlane, 2003).
This basic idea of the biological trait-based approach, first introduced
in terrestrial systems (Díaz and Cabido, 2001) and later adopted for
freshwater and marine systems (Bremner et al., 2003; Usseglio-Polatera
et al., 2000), was to move away from a taxonomic approach of community
analysis to a more meaningful functional approach.{~}

Thus far, the majority of trait-based studies investigate the effect of
anthropogenic stressors at a community level aiming to gain insights to
how trawling changes the traits, or to some extend functional
composition, of a community. Within this approach, traits are weighted
by the abundance or biomass of all species exhibiting the selected trait
(Bremner et al., 2003) and this pooled data is subsequently related to a
stressor such as trawling (Villéger et al., 2010, Hiddink et al., 2019).
The challenge with this approach is that the observed trait responses
cannot indisputably be linked to the stressor nor can the results of
such studies be easily generalized. This stems from the fact that the
individually analysed traits are in fact the result of a combination of
interdepend traits exhibited by the species. Some of the traits
expressed by a species may facilitate a certain response to a stressor,
while others may impede it. Thus, it is the interplay or sum of opposing
or synergistic traits that will determine the response of a species to a
stressor. As an example, species living on the surface of the seabed are
highly likely to be impacted by trawling (Tiano et al., 2020), however,
if surface dwelling species have a highly resistant shell and have a
large reproductive potential, they may survive trawling impacts and
quickly compensate for individual losses at a population level (Bremner
et al., 2005). If we had several species with a similar traits
combination dominating the community, it could wrongly be concluded that
trawling had little or no effect on organisms that inhabit the seabed
surface. While this may be true for the particular area analysed due its
species composition other areas may show different responses. Thus,
analysing the responses of traits at community level has the potential
to introduce bias and lead to spurious and inconsistent conclusions
about the impact of a stressor.{~}

To overcome some of these shortcomings, it has been suggested to group
species into vulnerability groups according to a set of traits that are
a-priori linked to a stressor or ecological function (De Juan et al.,
2014) .Within the present paper, we introduce new trait-based indexes
for benthic species and link this to a clearly defined ecosystem
function, i.e., bioturbation. The index is based on the observations
that the vulnerability of a species to trawling is not random but
related to specific morphological, life history as well as behavioural
characteristics (Jørgensen et al., 2016). In general, large, fragile and
slow reproducing species living on the surface of the seabed appear to
be the most affected by chronic trawling, while robust, small and fast
reproducing species tend to be the least affected (de Juan et al., 2012;
Jennings and Kaiser, 1998a; Van Denderen et al., 2015). From this
general observation, two subgroups of traits can be delineated, those
related to the physical resistance potential of a species (i.e. traits
related to body size, living habit, body form) and those related to its
recovery potential (i.e. traits such as small body size and fast
reproduction). Both of these trait groups contribute to the
vulnerability (or the opposite resilience) of species to trawling. The
development of our indexes followed the same logic and, therefore, we
first constructed two sub-indexes a) a physical ``Resistance index''
(RI), considering morphological and behavioural aspects (e.g., body
structure or living position), and b) a ``Recovery Potential index''
(RPI), considering traits related to the reproductive strategies and
population growth potential. These two indexes were subsequently
combined into an overarching index that we named ``\textbf{R}esistance
and \textbf{R}ecovery potential \textbf{I}ndex'' or RRI. This index can
on the one hand be used as a standalone index to attain a measure of the
vulnerability or resilience of a community to trawling, on the other
hand it can be used to explore the link between vulnerability and
ecosystem functions. Besides introducing the RRI and its sub-indices,
the present study aimed to validate and demonstrate the mulitple uses of
the index by applying it to benthic data from the north-eastern Irish
Sea \emph{Nephrops} fishing ground (Hinz et al., 2009) and linking it to
a well-established functional index, the bioturbation potential index
(BPI) developed by Queirós et al. (2013).{~}

This new trait-based index and approach provides a platform for a more
in-depth exploration of the effects of trawling on vulnerability and
ecosystem function, simultaneously. While the present study focuses on
one particular stressor and function, its simplicity could serve as a
template easily adapted and applied to other stressors and functions,
e.g., climate change (stressor), habitat provision (function), or
similar.{~}

\textbf{2. Methods}

Within this this section we first describe the calculation of the
indices, then outline the validation of these based on case study data
from the Irish Sea. For the validation, we investigate the effectivity
of the RRI to represent vulnerability to trawling based on observed
responses to trawling at a species level. Furthermore, we describe the
application of the RRI on a community level. Finally, we portray how we
link the vulnerability of species to function through the RRI using the
bioturbation potential index and how this can be used within scenario
simulations in data poor areas.{~}

\textbf{2.1 Calculations of indices}

To determine the vulnerability of species to trawling, we developed two
additive indices: the resistance index (RI) and the recovery potential
index (RPI). The combination of the two indices into a third index:
Resistance and Recovery Potential Index (RRI), that determine the
potential vulnerability of a species. All three indices are designed as
weighted directional indices where the final calculated score of a
species in the respective index reflects the contribution of all
relevant traits combined towards the objective of the RRI index, that is
expressing the vulnerability/resilience of species to trawling.{~}

\emph{2.1.1. RI and RPI calculations}

The rational for the RI and RPI indices follows from the objective to
describe the physical resistance and the reproductive potential of a
species to trawling. For this, we created simple indices using readily
available traits information, avoiding traits with known gaps. The
traits related to resistance included: body form, body texture, size and
environmental position. While traits related to the potential recovery
after disturbance were related to reproduction and growth: size, adult
longevity, reproductive frequency, development type, regeneration of
body parts, scavenger (see Table 1 and 2 for trait categories and their
rational). The single feeding type scavenger, was included as trawling
is known to significantly benefit species with this feeding mode
(Groenewold and Fonds, 2000; Tillin et al., 2006). The trait body size
was used in the two indices as size is related to the physical
resistance of a species to trawling, with larger species having a higher
tendency to be caught or damaged, and also to the potential to recover
after trawling, as small organisms tend to have faster growth and
reproductive cycles compared to large slower growing organisms with less
frequent reproduction (Jennings and Kaiser, 1998b).

Following the methodology in biological trait analysis, (Bremner et al.
2003, Törnroos \& Bonsdorff 2012) the traits linked to the different
aspects of the morphology, reproduction and behavior, used in the
indices, are collected for each species and compiled in a matrix. Trait
categories are then assigned to the selected traits; in our case study,
body texture had the following five categories: brittle, unprotected
soft tissue, thin exoskeleton or shell, durable /flexible, hard
exoskeleton or shell. Species are then scored for their affinity to a
trait category following the fuzzy scoring method using a scale of 0 (no
affinity) to 1 (high affinity), with a total score of 1 for each trait
(Bremner et al., 2003). The fuzzy coding allowed the species to vary in
the degree in which it exhibited affinity to a specific category within
a trait. The traits' assignment was based on available literature,
information from online databases (e.g. BIOTIC, www.marlin.ac.uk/biotic
and others see Appendix 1) and experts' knowledge. When no information
on a trait was available for a species, information for the genera was
considered; a minor proportion of cases had family or higher group level
information. All literature sources of trait information have been
acknowledged as well as the amount of expert knowledge in populating the
traits matrix used in this project (see S1).

To calculate the indices for a particular species, the fuzzy coding of
traits as described above was multiplied by the directional weighting
scores of the index. Scores were given on the logical proneness of a
specific trait category to be impacted by trawling; e.g., in the case of
``body form'', to be caught, broken or uprooted by a passing trawl,
i.e., they were ranked 1 to 3 respectively, with erect contributing
least to resistance and species with a low vertical depth contributing
the most. To ensure each trait category had the same influence on the
final index when calculating summed index scores, the trait category
rank values were multiplied by the maximum number of trait categories in
any one trait (which was 5) divided by the number of traits categories
in the observed trait (Balancing Score see table 1 and 2). The
contribution of each trait to the respective index was kept equal as
there is uncertainty about the precise strength of influence of each
individual trait towards the indices objectives, i.e., resistance or
recovery potential (for a summary of the directional weighting scores
and its rational see Table 1 and 2){ }.

\begin{center}
	\includegraphics[width=1\linewidth]{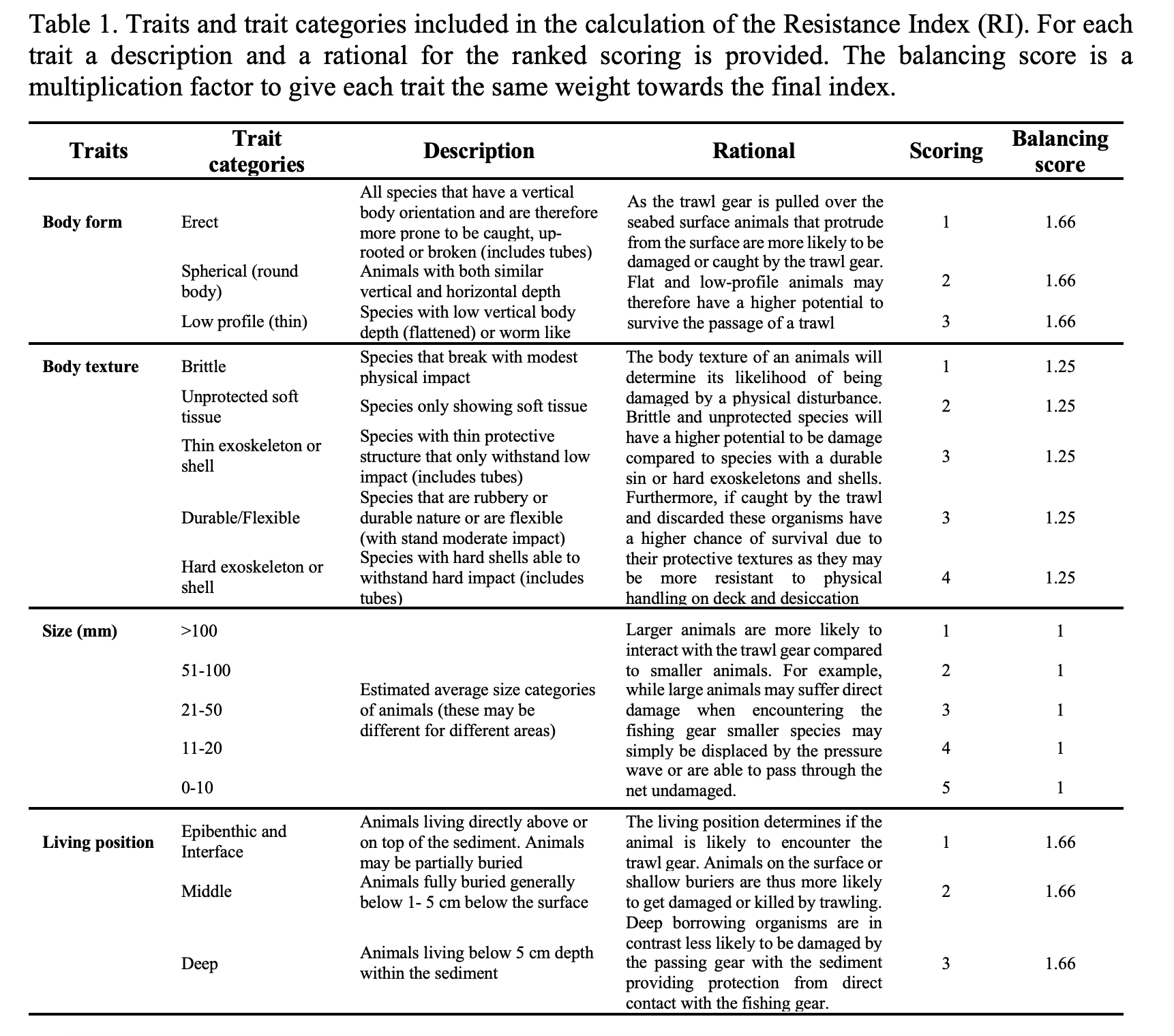}
	\label{table:table1}
\end{center}

\begin{center}
	\includegraphics[width=1\linewidth]{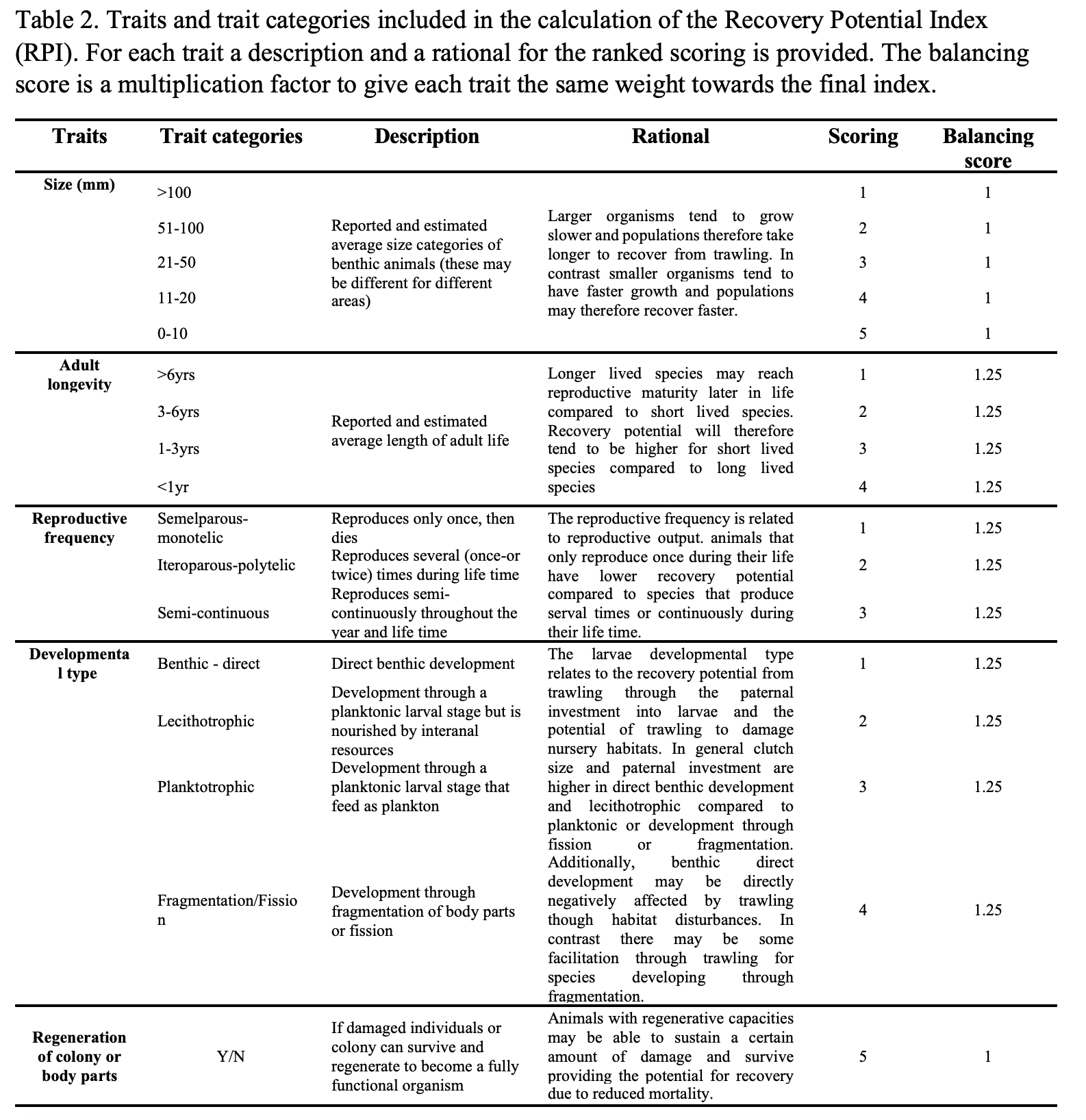}
	\label{table:table2}
\end{center}

To normalize both index scores, and thus providing results on a similar
scale (0 to 1), the following formula was used:

Values close to 1 for the RI of a species indicates that it is
potentially highly resistant to trawling, while a value of 1 for the RPI
indicates high recovery potential due to the associated life history
traits.{~}

\emph{2.1.2. RRI calculations}

To calculate the Resistance and Reproductive Potential Index (RRI), the
mean of both individual normalized index scores (RI and RPI) was
calculated (from 0= highly vulnerable, to 1= highly resilient). The RRI
species scores were subsequently categorised into five levels to
identify species with similar scores for the subsequent analyses: 0.8-1
= Very High RRI, 0.6-0.79 = High RRI, 0.4-5.9 = Moderate RRI, 0.2-4.9 =
Low RRI and 0-0.19 =Very Low RRI.{~}

\textbf{2.2.Validation of the RRI index}

\emph{2.2.1. The Irish Sea case study and macrofauna data}

To validate the developed indices, we used macrofauna data collected
over an active fishing ground for Norway lobster (\emph{Nephrops
norvegicus}, Linnaeus) and gadoid fish in the north-eastern Irish Sea.
This data was previously published investigating the effects of trawling
on a taxonomic level by Hinz et al. (2009). For more detailed
information about the sampling design, as well as the fishing effort
calculation, refer to Hinz et al. (2009). Macrofauna data was collected
at 15 sites over a gradient of fishing intensity varying from 1.3 times
trawled/year to 18.2 times trawled/year. Both benthic infauna (day grab
samples 0.1 m\textsuperscript{2}) and epifauna (2-m beam trawl) were
sampled and standardized to m\textsuperscript{2} biomass. The two
datasets were combined into one single data matrix for the subsequent
analyses (for benthic data see S2 and for trawling intensity and
physical characteristics of sampling sites S3). The methodology for
calculating the two sub-indices and the final RRI index describe above
was applied to the macrofauna dataset.

\emph{2.2.2 Effectivity of the RRI to represent vulnerability to
trawling on a species level{~}}

The effectivity of the indices to represent the overall vulnerability of
species to trawling was validated by comparing the calculated RRI scores
of individual species to observed responses of those same species over a
gradient of trawling intensity (as recorded by Hinz et al., 2008), using
linear regression slopes. Prior to analysis, the individual species
biomass data was normalized. For the validation analysis, we considered
only species that had sufficient data: occurred at least over 4 stations
out of 15, i.e., over 30\%. We expected that for our RRI index to be
valid, there should be a significant correlation between the species RRI
scores and the individual species regression coefficients from the
observed data. The correlation between the RRI index and regression
coefficients was calculated for species groups with different
occurrences over the sampling stations. Here we wanted to see if the
correlations would improve when using data of species that had
successively higher occurrences over our sampling area. The assumption
made was that the responses of common species would contain less errors
related to false zero observations (i.e., due to absence at sites
unrelated to trawling) making the correlation between RRI score and
coefficient more robust and representative for the validation of our
index. However, considering that some of the less common species are
also those highly vulnerable to trawling, we decided to present the
results of all correlations above the aforementioned minimum threshold
of 30\% occurrence over the sampling stations (see above). Three species
in the validation analysis form a commensal type of association with
another larger species. The small bivalve \emph{Tellimya ferruginosa}
lives associated with irregular urchins such as \emph{Echinocardium}
spp, while the small bivalve \emph{Kurtiella bidentata} and the
polychaeta \emph{Podarkeopsis helgolandicus} are associated with brittle
stars such as \emph{Amphiura} spp\emph{.} In these cases, the lower
scoring host species RRI was used and not the original score calculated
based on the species traits. The assumption was that the hosts' response
to trawling would have a greater influence on the response of the
associated species than the calculated species RRI score.

\emph{2.2.3. Modelled responses based on RRI grouping of species}

We investigated the observed responses of species, when grouped
according to their RRI index scores, test expected species responses to
trawling (i.e., Low RRI species showing a strongly negative trend,
followed by a less negative response for species of the Medium RRI group
and non or a positive response for High RRI species). RRI group
responses were modelled on a species level using generalized mixed
modelling (GLMM) on the observed case study data, using the RRI group
category as random effect (i.e. a grouping factor). As above, we used
normalized biomass data versus fishing effort to calculate the
regression coefficients by species. The resulting model relationships
are thus the mean responses of species belonging to a RRI group. For the
analysis, we pooled species with very low and low RRI scores as the
former group only contained two species. Species with very high RRI
scores did not exist within our study area, and therefore we
investigated the response of three RRI groups: Low, Moderate and High.
These three groups were maintained for all subsequent analysis.{ }Note
that the RRI group High only contained three species. The GLMM models
were tested for their significance following procedures outlined by Zuur
et al.(2009)

\emph{2.2.4. Relationship of indices (RI, RPI, RRI) to trawling on a
community level}

The relationship between scores of the indices at a community level and
trawling were explored by linear regression models. An average score of
the respective index on a community level were calculated by multiplying
species index scores with species biomass at a respective station,
subsequently summing all individual scores of that station and dividing
this sum by the total biomass of the respective station.

\textbf{2.3. Linking vulnerability to function}

\emph{2.3.1. Response of bioturbation potential (BPc) to trawling}

The Bioturbation Potential Index (BPc) introduced by Queirós et al.
(2012) was selected to demonstrate the link between RRI and ecosystem
functions.

BPc takes into account the average size and abundance of organisms
attained from sample data and combines these with bioturbation weighting
factors based on categorical scales describing the mobility and sediment
reworking of an organism (Queirós et al 2012). The following formula
describes the calculation of BPI of a benthic community:

$ BPc=\sum_{i=1}^{n}{\sqrt{\frac{\mbox{Bi}}{\mbox{Ai}}\; x\mbox{Ai}\; x\; Mi\; x\; Ri}} $

Bi and Ai are the biomass and abundance of species/taxon \emph{i} in a
sample. Mobility (Mi) range from 1 (living in a fixed tube) to 4 (free
three-dimensional movement via burrow system). Sediment reworking (Ri)
from 1 (epifauna that bioturbates at the sediment--water interface) to 5
(regenerators that excavate holes, transferring sediment at depth to the
surface). For the present paper we used the trait information provided
by Queirós et al (2012) for 1033 macrofaunal species in the case study
data by extracting the relevant information to match our species list.
Most species were already included within the database. Only 7 species
were not found in the database provided by Queirós et al (2012) and
information for these species was therefore added by the present study
(see S4 for a full of species list and their relevant biological traits
categorisation) using published descriptions of species bioturbation
behaviour, or in its absence, information on closely related species.

Table 3. Summary of trait categories and scores used to calculate BPc as
described by Queirós et al (2012)

\begin{longtable}[]{@{}llll@{}}
\toprule
\begin{minipage}[t]{0.22\columnwidth}\raggedright
\textbf{Mobility (Mi)}\strut
\end{minipage} & \begin{minipage}[t]{0.22\columnwidth}\raggedright
\textbf{Score}\strut
\end{minipage} & \begin{minipage}[t]{0.22\columnwidth}\raggedright
\textbf{Sediment reworking types (Ri)}\strut
\end{minipage} & \begin{minipage}[t]{0.22\columnwidth}\raggedright
\textbf{Score}\strut
\end{minipage}\tabularnewline
\toprule
\begin{minipage}[t]{0.22\columnwidth}\raggedright
Organisms that live in fixed tubes\strut
\end{minipage} & \begin{minipage}[t]{0.22\columnwidth}\raggedright
1\strut
\end{minipage} & \begin{minipage}[t]{0.22\columnwidth}\raggedright
Epifauna\strut
\end{minipage} & \begin{minipage}[t]{0.22\columnwidth}\raggedright
1\strut
\end{minipage}\tabularnewline
\begin{minipage}[t]{0.22\columnwidth}\raggedright
Organisms with indicates limited movement\strut
\end{minipage} & \begin{minipage}[t]{0.22\columnwidth}\raggedright
2\strut
\end{minipage} & \begin{minipage}[t]{0.22\columnwidth}\raggedright
Surficial modifier\strut
\end{minipage} & \begin{minipage}[t]{0.22\columnwidth}\raggedright
2\strut
\end{minipage}\tabularnewline
\begin{minipage}[t]{0.22\columnwidth}\raggedright
Organisms with slow, free movement through the sediment matrix\strut
\end{minipage} & \begin{minipage}[t]{0.22\columnwidth}\raggedright
3\strut
\end{minipage} & \begin{minipage}[t]{0.22\columnwidth}\raggedright
Upward and downward conveyors\strut
\end{minipage} & \begin{minipage}[t]{0.22\columnwidth}\raggedright
3\strut
\end{minipage}\tabularnewline
\begin{minipage}[t]{0.22\columnwidth}\raggedright
Organisms with free movement, that is, via burrow system\strut
\end{minipage} & \begin{minipage}[t]{0.22\columnwidth}\raggedright
4\strut
\end{minipage} & \begin{minipage}[t]{0.22\columnwidth}\raggedright
Biodiffusors\strut
\end{minipage} & \begin{minipage}[t]{0.22\columnwidth}\raggedright
4\strut
\end{minipage}\tabularnewline
\begin{minipage}[t]{0.22\columnwidth}\raggedright
\strut
\end{minipage} & \begin{minipage}[t]{0.22\columnwidth}\raggedright
\strut
\end{minipage} & \begin{minipage}[t]{0.22\columnwidth}\raggedright
Regenerators\strut
\end{minipage} & \begin{minipage}[t]{0.22\columnwidth}\raggedright
5\strut
\end{minipage}\tabularnewline
\bottomrule
\end{longtable}

The relationship between trawling versus BPc for different RRI groups
(i.e. low, medium, high) was investigated using linear regression
models. At each station the summed bioturbation potential per station
was calculated for each group. To linearize the data all regressions
were performed after log transformation.

\emph{2.3.2. Scenario modelling of the effect of trawling on
bioturbation}

We undertook scenario or stress test modelling emulating a situation
where little benthic data is available but the potential consequences of
species removal or reductions of abundance are to be explored. For the
scenario modelling, the data from the least impacted site in the Irish
Sea case study was used as the baseline. We modelled the effect on
bioturbation potential from 1) directed elimination and 2) reductions in
species abundances, based on the species' RRI ranking, which provided an
indication of the species' vulnerability to trawling.{~}

 \emph{2.3.3. Directed extinction scenario of low RRI or high RRI species}

We simulated the complete extinction of the 5, 10, 20, 30 and 40 percent
of the most vulnerable species based on their RRI ranking (low RRI
scores). We contrasted this deletion of species with random deletions,
i.e., removing species at random from the species list considering the
same number of species as removed for the low RRI ranking species.
Random removals were performed by the random subsampling function in R
(sample, base v3.6.2) and repeated 999 times. The resulting ``directed
deletion'' scenario can be compared to the position and slopes of the
``random deletion''. If the slope of the directed removal is found to be
above the random slope, vulnerable species are not strongly linked to
that function. In contrast, if the slope is found below the random
deletion slope, there is a strong indication that some of the vulnerable
species contribute disproportionality to the function analysed. If the
slopes of random and directed deletion are similar, vulnerability and
function are not linked.{~}

Furthermore, we reversed the removal to assess the contribution of
non-vulnerable species to trawling, i.e., species ranked with a high RRI
score to the bioturbation potential. In this scenario, we removed 5, 10,
20, 30 and 40 percent of the most resistant and resilient species from
the species list and subsequently calculated the BPc. All scenario
responses were analysed through linear regression models.

\emph{2.3.4.Reduction in abundance of different RRI groups{~}}

Within a set of secondary scenarios, we demonstrate the effect of
reducing the abundance of species belonging to the three RRI groups on
the BPc. Abundances were reduced from 10 to 90\% for a certain RRI
group, while species within other RRI groups were not reduced in
abundance for community BPc calculations. Furthermore, we created a
random group by selecting species at random from the species list and
considering the number of species found within the low RRI group as a
reference. Random abundance reductions were performed as described for
the extinction scenarios. The impact of abundance reductions of
different vulnerability groups (RRI) on BPc can be explored by comparing
the slopes and their relative position.{ }All scenario responses were
analysed through linear regression models.

\textbf{3. Results}

\textbf{3.1. Validation of the RRI index}

\emph{3.1.1. Effectivity of the RRI to represent vulnerability to
trawling on a species level{~}}

To confirm the assumption that species with a low RRI should respond
more strongly to trawling compared to higher RRI species, we correlated
individual RRI scores of species with their observed regression
coefficients. Considering all 54 species, we found a 0.22
r\textsuperscript{2} correlation (Figure 1A). When considering species
with subsequently higher frequency of occurrence over the study area,
the r\textsuperscript{2} values increased up to 0.61 (Figure 1B) for the
12 species that occurred at all sampling stations.\emph{{~}}

\emph{}\\

\begin{figure}
	\centering
	\includegraphics[width=1\linewidth]{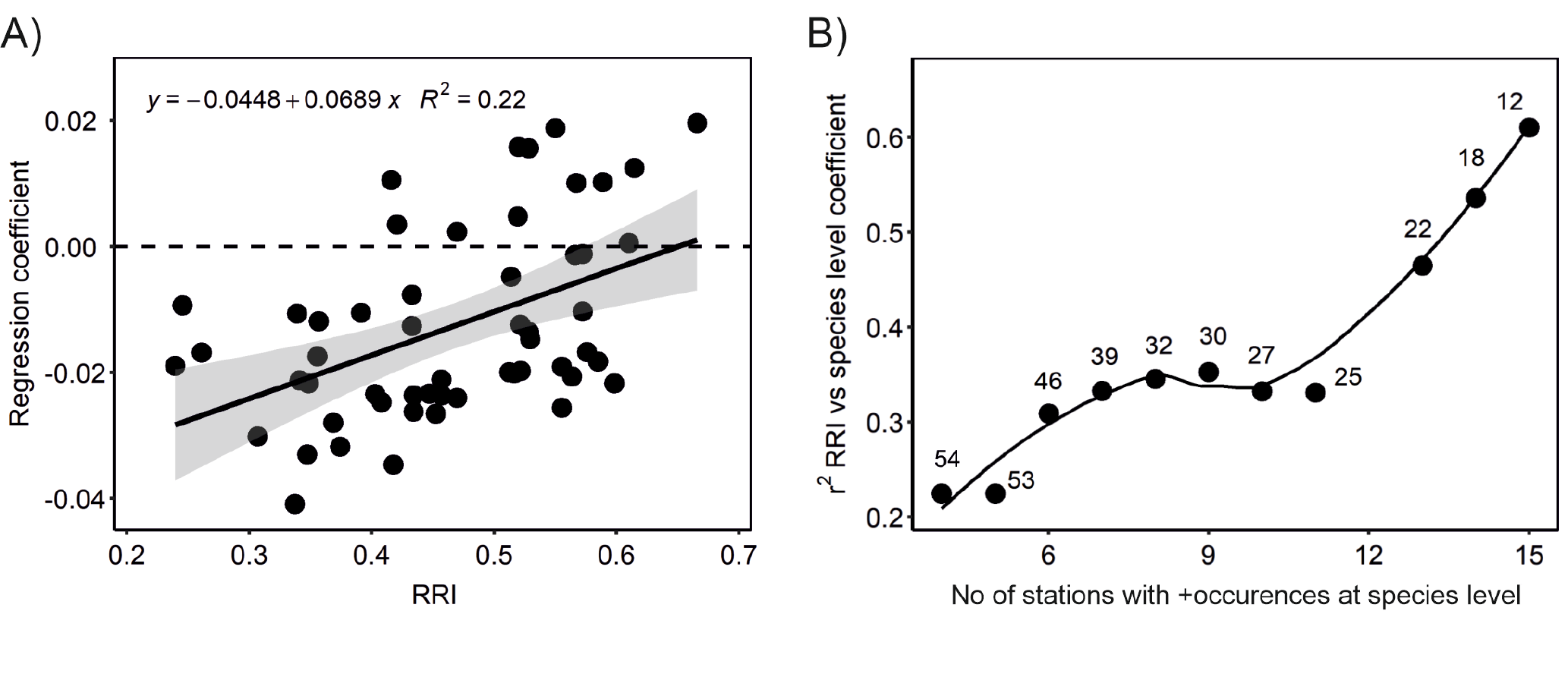}
	\caption{A) Relationship between the RRI scores of benthic species and
		their regression coefficients from the observed relationships between
		trawling intensity and normalized biomass. B) Trend in the
		r\textsuperscript{2} of the relationship featured in 2A consecutively
		excluding species that had many zero observations. The number above each
		data point indicates the number of species at that occurrence level,
		e.g., at the extreme right 12 species occurred at all 15 stations
		sampled and the relationship between RRI and coefficient had an
		r\textsuperscript{2} of 0.61 (for regression statistics see S5)}
	\label{fig:figure1}
\end{figure}

\emph{3.1.2. Modelled responses to trawling based on RRI grouping of
species}

The estimated mean response of species belonging to the three RRI groups
using GLMM showed that species belonging to the Low RRI group, i.e.,
species with a low resistance and reproductive potential, mostly had a
negative response to increases in trawling intensities (Figure 2). This
was followed by a slightly less negatively sloped relationship for the
Moderate RRI species and a positive relationship for the High RRI
species. The relationship for Low and Moderate RRI species was found to
be statistically significant, while the estimated relationship for High
RRI species was not (Table of model statistics see S6, figure showing
individual and mean responses of grouped species S7).{~}

\emph{}\\

\begin{figure}
	\centering
	\includegraphics[width=0.9\linewidth]{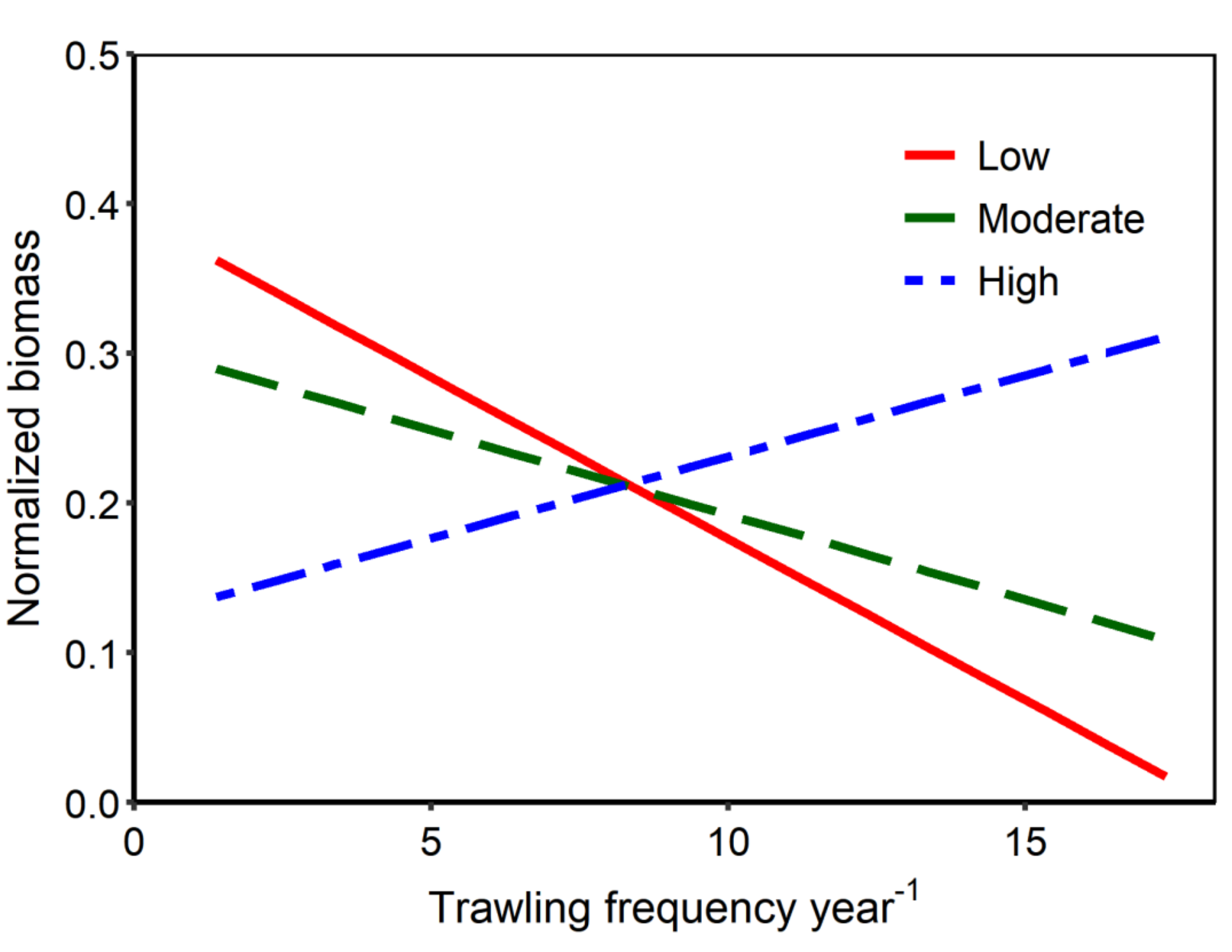}
	\caption{Modelled mean responses of species belonging to the three RRI groups (Low; Moderate and High) using GLMM. The regression slopes were statistically significant for Low and Moderate RRI groups of species, while not for the High RRI group (see S6 for a detailed result table). Note that the High RRI group only contained three species.}
	\label{fig:figure2}
\end{figure}

\emph{3.1.3 Relationship of indices (RI, RPI, RRI) with trawling
intensity on a community level}

The individual species index scores, when calculated across the entire
community, showed significant increases (p \textless{} 0.05) in all
three indices in response to trawling (Figure 3 A-C). This indicates
that the benthic communities in the case study area were increasingly
composed of species with high resistance and reproductive potential.{~}

\begin{figure}
	\centering
	\includegraphics[width=1\linewidth]{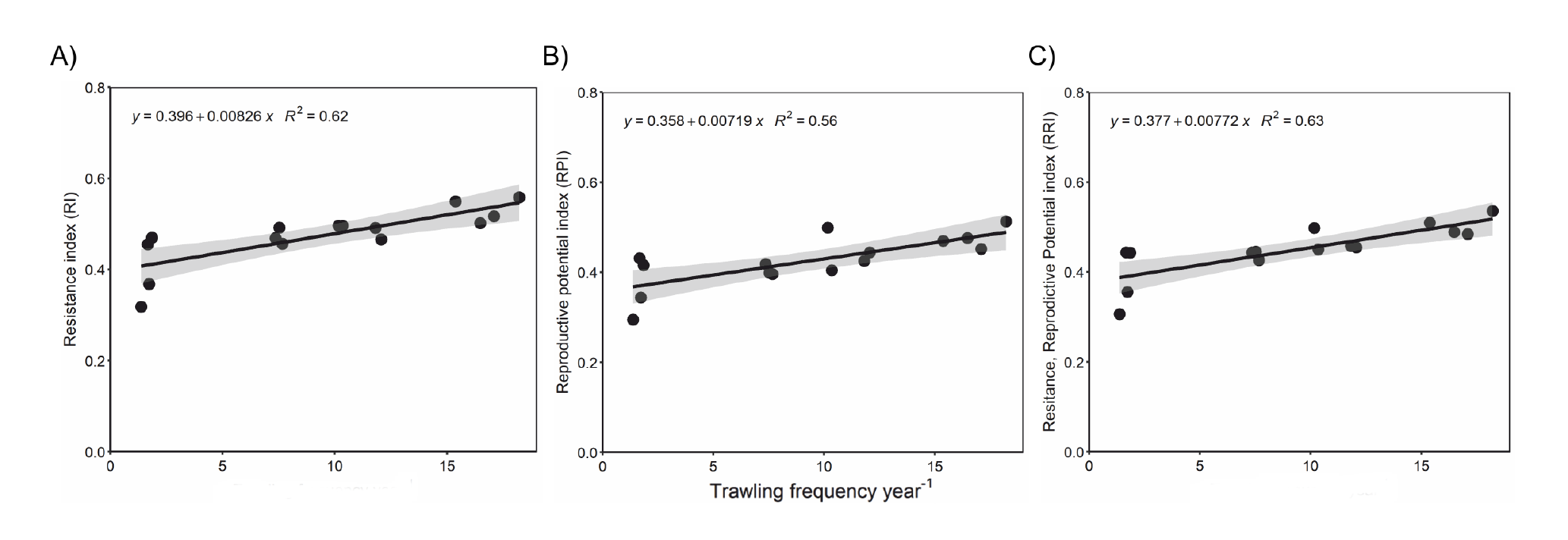}
	\caption{Relationship between trawling intensity, as times swept per annum, and the three indicators A) physical Resistance Index (RI), B) the Reproductive potential index (RPI) and C) the combined Resistance, Reproductive potential Index (RRI).}
	\label{fig:figure3}
\end{figure}

\textbf{3.2. Linking vulnerability to function{~}}

\emph{Response of bioturbation potential (BPc) to trawling intensity}

The BPc responded overall negative to increased intensities of trawling
in the case study area (Figure 4a). When considering the different RRI
groups, a strong negative relationship of the BPc was found for the low
RRI group species (i.e. highly vulnerable species), while BPc for
moderate RRI species responded negatively, but with a less strong slope
(Figure 4b, S6). High RRI species, in contrast, increased their BPc over
the trawling gradient.

\textbf{}\\

\begin{figure}
	\centering
	\includegraphics[width=1\linewidth]{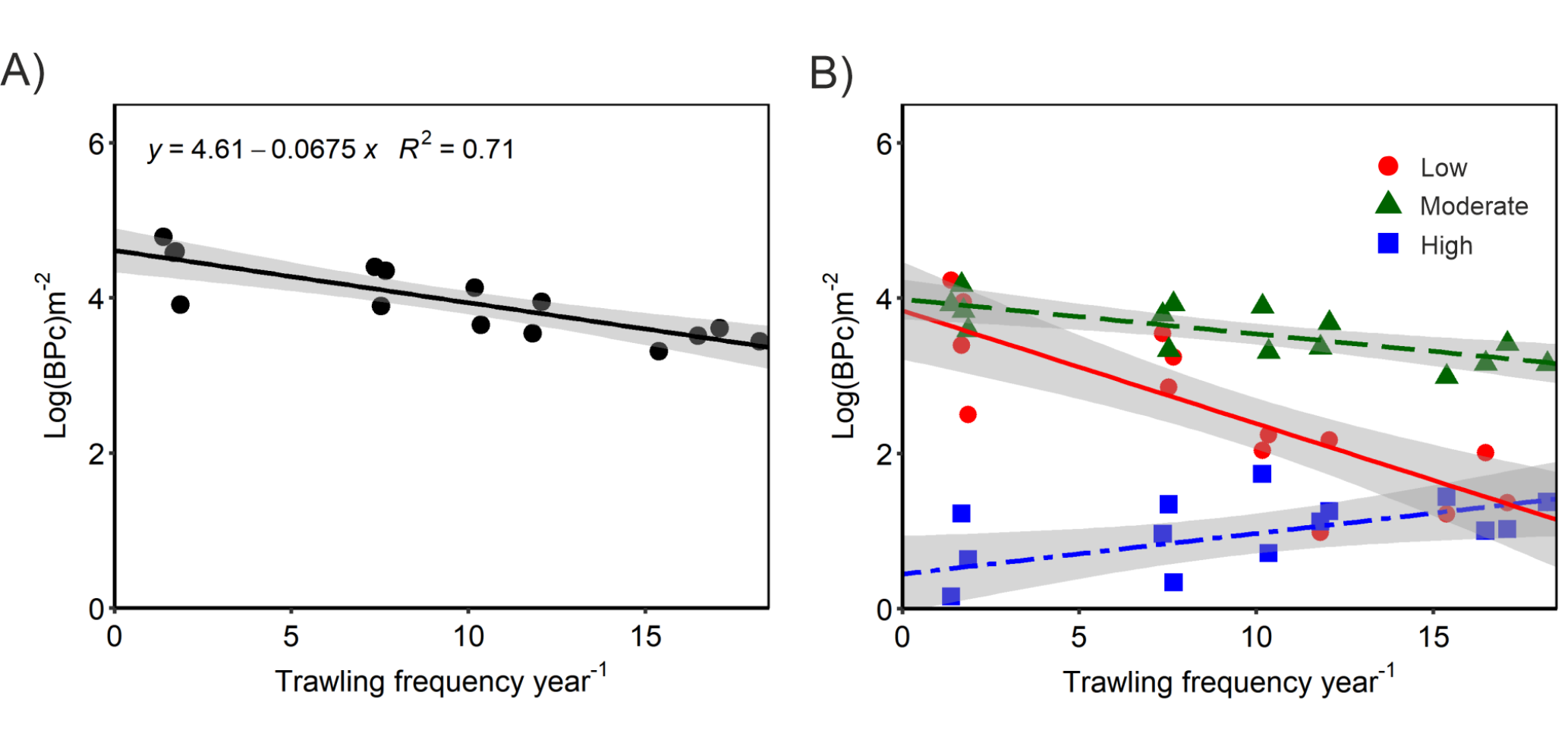}
	\caption{A) Response of the log transformed bioturbation potential (BPc) with increasing trawling intensity as times trawled per annum. B) Responses of the log transformed bioturbation potential (BPc) for the three RRI groups. For more details about the regression results the reader is referred to S6.}
	\label{fig:figure4}
\end{figure}

\textbf{3.4. Scenario modelling of the linkage between RRI and function}

\emph{3.4.1. Directed extinction scenario of a low RRI or a high RRI
species{~}}

The simulation of removing species from the station least affected by
trawling in the order of vulnerability, from lower to higher RRI scores
in 10\% steps, showed a strong negative effect on the bioturbation
potential (Figure 5A). Reversing the removal of species from the
community, from higher to lower RRI species, showed a less steep
negative response. The simulated mean random removal of species
demonstrated an intermediate negative response with the slope being
located between the former two (Figure 5 A, S6).

\begin{figure}
	\centering
	\includegraphics[width=0.9\linewidth]{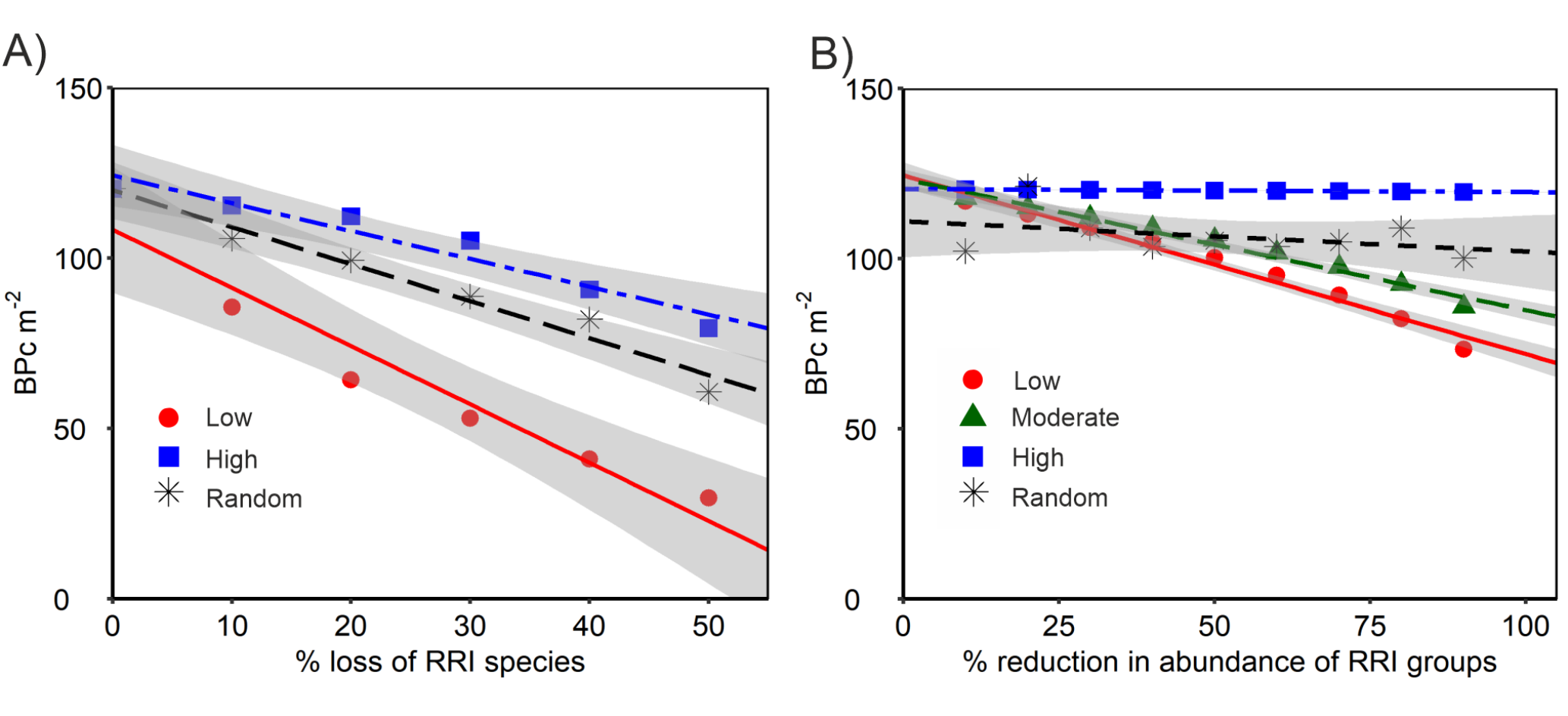}
	\caption{Simulation scenarios taking the bioturbation potential (BPc) of the least impacted site as a baseline. A) Consecutive elimination of species from low to higher RRI species (Low, red solid line) and from high to lower RRI species (High, blue double dashed line), i.e., removal of 10-50\% of the species. The grey dashed line (stars) represents the removal of species in a random order (mean of 999 permutations). Note that this graph is not referring to RRI groups. B) Scenario of reducing the abundance of different RRI groups on the bioturbation potential (BPc), from 10 to 90\%, while keeping the contributions from the other groups constant. For the random response, we reduced the abundances of randomly selected species using the same number of species as in the low RRI group. The random reductions represent mean values of 999 permutations.}
	\label{fig:figure5}
\end{figure}

\emph{3.4.2. Reduction in abundance of RRI groups and its effect on
bioturbation{~}}

Within this scenario, the effect of reducing the abundance of different
RRI groups by 10 to 90\% and its effect on BPc was investigated. The
abundance of respective groups was lowered, while keeping the
contributions from the other groups constant, providing an estimate of
how the different levels of reduction affected community BPc. There was
a strong negative response in BPc when the abundance of species with low
(vulnerable) RRI scores was reduced (Figure 5B). The reduction in
abundance of the moderate RRI had a similar negative effect with a
slightly less steep slope (Figure 5B). The reduction in abundance of the
high RRI group, as well as the reduction of random species, did not
lower the BPc to the same extent as the other two groups (Figure 5B, see
S6 for model statistics).{~}

\textbf{4. Discussion}

\textbf{4.1. Why a new trait-based approach for fishing impact studies?}

The impetus of creating a benthic resistance and recovery potential
index (RRI) was based on the simple realization that a species
represents a combination of traits that are interdependent of each
other. It is the interplay of a diversity of traits, with opposing or
synergistic effects, that determine the response of a species to an
external impact. The introduced RRI index aims to pay tribute to the
fact that trawling impacts ultimately occur at a species level and that
a better understanding at this level would allow us to make better
predictions on benthic communities' responses to trawling. Equally, as
demonstrated within this study, understanding the vulnerability on a
species level can allow us to model or stress test the response of
species, communities and functions to different pressure scenarios,
e.g., by removing or reducing the most vulnerable species and assessing
the consequences for ecosystem functions, like bioturbation in the
present study.{~}

Another realisation that furthered the development of this index was
that there are certain traits that are more directly linked to a
stressor, while other traits respond only due to their association to
those more directly linked traits. For example, a trait like life span
could be regarded as indirectly related to the physical impact of the
trawl gear, while living position, body texture or size are traits more
closely linked to how a species will be directly affected by the
trawling gear. We therefore created additive indices, combining traits
that reflected a directional trend towards the resistance and
reproductive potential of a species. This type of additive index, where
various traits are combined to reflect a particular response or
function, has been used in other traits-based studies as for example in
the bioturbation potential index introduced by Queirós et al. (2013). We
believe that the development of this type of index has great potential
and could be refined for other fishing impacts, habitat types, as well
as for other human impacts or functions in general.{~}

Similar to other trait-based indices, there is an element of
subjectivity over which traits to integrate into the indices presented
in this study. In this case, the traits were chosen based on the
currently best available knowledge and mechanistic understanding about
trawling impacts on muddy habitats and the ease of attaining such data
for most species. It is though crucial to provide transparency over the
index construction and evaluate its performance in a validation process.
In this study we tried to address both of these aspects by comparing its
performance to real observational data and by providing the required
references of our trait data including the amount of expert judgement
that was used to populate the underlying traits matrix (see S1).

\textbf{4.2. Validation of the Resistance and Recovery Potential
Index{~}}

The RRI index was validated by estimating the mean observed response of
different RRI groupings of species to trawling using data previously
published by Hinz et al. (2008). The mean response of species grouped
after their vulnerability (RRI score) demonstrated the predicted pattern
based on theory about trawling impacts on benthos (Jennings and Kaiser,
1998b). Thus, vulnerable species (Low RRI) respond strongly with a
negative trend to trawling pressure, followed by moderately vulnerable
species (Moderate RRI), and species with a high resistance and
reproductive potential (High RRI) showing no significant response.
Similarly, when comparing the individual slopes of observed species
responses with their respective RRI index scores, we found a significant
correlation between the two. The results of this part of the validation
analysis provided confidence that the traits chosen for our indices and
their scoring reflected the relative vulnerability or resilience of
benthic species to trawling intensities. Nevertheless, the difference in
fits of the correlations observed, depending on which species were
included in the relationship, also revealed the predicaments over the
biases of this type of validation. Species with rare occurrences over
the trawling gradient often had many zero observations introducing
uncertainty over estimated species responses (slopes) to trawling.
Therefore, species with very few observations were removed from the
analysis. However, these species may have been rare due to the trawling
impact and thus we removed species strongly related to trawling impacts.
For example, the sea pen \emph{Virgularia mirabilis,} a highly
vulnerable species with one of the lowest RRI scores recorded within our
study, only occurred at the two least fished sites and thus could not be
included in the validation analysis. Contrary, other species included in
the analysis showed trends based on few data points that were
statistically not significant, potentially having introduced
considerable bias into the evaluation analysis. Focusing the evaluation
analysis only on the common species that occurred at most sites appeared
to show the best fit between the observed regression slopes and our RRI
index. Focusing the entire evaluation only on these species, however,
would have reduced the generality of the analysis and would not have
allowed the estimation of vulnerability (RRI) group responses as
described at the beginning of the paragraph.{~}

While the results of the validation analyses are encouraging, some
details over the calculation of the RRI index need to be appreciated, in
particular with respect to the scoring and weighting of the index and
sub-indices. We choose to give each trait within each sub-index, the
physical Resistance Index (RI) and the Reproductive potential index
(RPI), roughly the same influence on the final score. It is quite likely
that some traits have a higher influence on the resilience or
reproductive potential of a species. However, currently there exists
limited understanding over the relative importance of individual traits
and their contribution towards a function. For example, in the case of
resistance, it is difficult to scale how much more important the trait
living position is compared to the body morphology. Equally, we gave
each of the two sub-indices the same weighting in calculating the final{
}Resistance Reproductive potential RRI index as we currently cannot
determine which component makes a species more resilient to chronic
trawling disturbance. As our understanding increases about trawling
impacts, future versions of the indices could be weighted reflecting new
emerging insights.{~}

\textbf{4.3. The application of the resistance and reproductive potential
  indices}

The indices introduced within this study can be used in several distinct
ways, a) they can be applied to investigate the changes in community
resistance and reproductive potential as a response to a trawling
gradient; b) they allow to explore the species contribution towards a
response in terms of community biomass but also function, by the
division of species into different resilient or vulnerability groups;
and c) the can be used in stress test scenarios, by eliminating species
or reducing their abundance/biomass depending on their vulnerability,
and exploring its effect on linked ecosystem functions. The application
of the indices to the fishing gradient study by Hinz et al. (2009)
demonstrated well the multiple uses of the developed indices. All three
indices (RI, RPI and RRI) showed significant positive responses on a
community level to increasing trawling intensities. Thus, trawling,
through the loss of more vulnerable species, increase the overall
resilience of benthic communities in the Irish Sea with respect to their
physical resistance and reproductive potential. While the term community
resilience may have positive connotations depending on the context
(Baggini et al., 2015; Knapp et al., 2001), in the present study it
conveys the opposite. This is in line with many other studies on
physical disturbances that found that communities in highly disturbed
locations, anthropogenically or naturally, hold more resilient species
adapted to this disturbance (de Juan et al., 2009; Sciberras et al.,
2013). Resilience can therefore not be judged as a positive community
aspect \emph{per se} and the term should therefore be used with care
with respect to the advocated preservation of resilience to affront
ecosystem change (Gladstone-Gallagher et al., 2019). In general, our aim
should be to strive to preserve functional resilience
(Gladstone-Gallagher et al., 2019) through the protection of vulnerable
species that make a considerable contribution through their abundance or
biomass to a function. In this respect, our study found a strong impact
of trawling on the bioturbation potential that experienced a loss of
81\% along the trawling gradient studied. By having subdivided the
species in vulnerability groups, we could explore the contribution of
each group to this functional loss. High and moderately vulnerable
species contributed with 55\% and 45\% respectively. Among those species
with a high contribution towards bioturbation were, for example, the
brittle star \emph{Amphiura} spp., the irregular urchin
\emph{Echinocardium} spp. and the polychaete \emph{Pectinaria auricoma}.
In contrast, high resistant or low vulnerable species increased their
biomass over the trawling gradient and compensated to a small extent the
loss of the two other groups. Without this compensation, total
bioturbation potential loss along the trawling gradient would have been
3\% higher. The results of this study demonstrate that a large part of
the bioturbation function was linked to vulnerable species and that
their removal had a considerable effect on the provision of this
function.

The scenario analysis performed on the least impacted site of the Irish
Sea case study demonstrated the strength of the index for a data poor
scenario, where only a few benthic samples are available. The simulation
of sequential species loss from both the most vulnerable to the highest
resilient, and vice versa, reflected well the trends identified in the
observed data, demonstrating that similar conclusions about the
importance of vulnerable species and their link to bioturbation could
have been drawn from sampling one single site.

\textbf{5. Conclusion}

We feel that despite the adolescent nature of our index, it is
sufficiently advanced to be discussed, applied and hopefully developed
further by the scientific community. The principle idea to move away
from descriptive community wide trait-based analysis towards a more
focused and mechanistic trait-based approach, which appreciates the
integrate nature of traits unified in a species, should enable
scientists to develop new approaches that will be more universal, when
it comes to impact studies, and potentially have higher predictive
power. With respect to the presented RRI index and its sub-indices, it
means that the vulnerability and the functional potential (e.g.,
bioturbation) of benthic communities can be determined and predicted for
almost any location provided we have the associated traits data for the
local species' pool. Using the traits of individual species in this way,
allows for stress testing communities through developing scenarios that
may reflect future environmental change or management decisions (e.g.,
changes in fishing effort regulations). Within our study, we presented
an index that was directed toward the physical impact of trawling, but
equally, indices could be developed for other fishing gears or stressors
such as climate change or eutrophication, combining relevant species
traits linked to the respective stressor. As the introduced indices are
conceptually simple, they should be easily adaptable to other scenarios
and situations. To conclude, we demonstrated that through our indices,
new insights into the link of species vulnerability and function in
regard to trawling impact can be gained. While further testing and
development of our index will be required, we hope that our approach
will lead to a new scientific discourse with respect to the trait-based
approach and its potential to increase our understanding of functional
ecology.

\textbf{References}:

Baggini, C., Issaris, Y., Salomidi, M., Hall-Spencer, J., 2015.
Herbivore diversity improves benthic community resilience to ocean
acidification. J. Exp. Mar. Bio. Ecol.
https://doi.org/10.1016/j.jembe.2015.04.019

Berg, T., Fürhaupter, K., Teixeira, H., Uusitalo, L., Zampoukas, N.,
2015. The Marine Strategy Framework Directive and the ecosystem-based
approach - pitfalls and solutions. Mar. Pollut. Bull.
https://doi.org/10.1016/j.marpolbul.2015.04.050

Biedron, I.S., Knuth, B.A., 2016. Toward shared understandings of
ecosystem-based fisheries management among fishery management councils
and stakeholders in the U.S. Mid-Atlantic and New England regions. Mar.
Policy. https://doi.org/10.1016/j.marpol.2016.04.010

Bremner, J., Frid, C.L.J., Rogers, S.I., 2005. Biological traits of the
North Sea benthos: does fishing affect benthic ecosystem function? Am.
Fish. Soc. Symp.

Bremner, J., Rogers, S.I., Frid, C.L.J., 2006. Methods for describing
ecological functioning of marine benthic assemblages using biological
traits analysis (BTA). Ecol. Indic. 6, 609--622.
https://doi.org/10.1016/j.ecolind.2005.08.026

Bremner, J., Rogers, S.I., Frid, C.L.J., 2003. Assessing functional
diversity in marine benthic ecosystems\,: a comparison of approaches
254, 11--25.

Clark, M.R., Althaus, F., Schlacher, T.A., Williams, A., Bowden, D.A.,
Rowden, A.A., 2016. The impacts of deep-sea fisheries on benthic
communities: A review. ICES J. Mar. Sci.
https://doi.org/10.1093/icesjms/fsv123

de Juan, S., Demestre, M., Thrush, S., 2009. Defining ecological
indicators of trawling disturbance when everywhere that can be fished is
fished: A Mediterranean case study. Mar. Policy.
https://doi.org/10.1016/j.marpol.2008.11.005

de Juan, S., Hewitt, J., Thrush, S., Freeman, D., 2015. Standardising
the assessment of Functional Integrity in benthic ecosystems. J. Sea
Res. 98, 33--41. https://doi.org/10.1016/j.seares.2014.06.001

de Juan, S., Hinz, H., Sartor, P., Vitale, S., Bentes, L., Bellido,
J.M., Musumeci, C., Massi, D., Gancitano, V., Demestre, M., 2020.
Vulnerability of Demersal Fish Assemblages to Trawling Activities: A
Traits-Based Index. Front. Mar. Sci.
https://doi.org/10.3389/fmars.2020.00044

de Juan, S., Moranta, J., Hinz, H., Barberá, C., Ojeda-Martinez, C.,
Oro, D., Ordines, F., Ólafsson, E., Demestre, M., Massutí, E., Lleonart,
J., 2012. A regional network of sustainable managed areas as the way
forward for the implementation of an Ecosystem-Based Fisheries
Management in the Mediterranean. Ocean Coast. Manag. 65, 51--58.
https://doi.org/10.1016/j.ocecoaman.2012.04.024

De Juan, S., Thrush, S.F., Hewitt, J.E., Halliday, J., Lohrer, A.M.,
2014. Cumulative degradation in estuaries: Contribution of individual
species to community recovery. Mar. Ecol. Prog. Ser.
https://doi.org/10.3354/meps10904

Díaz, S., Cabido, M., 2001. Vive la différence: Plant functional
diversity matters to ecosystem processes. Trends Ecol. Evol.
https://doi.org/10.1016/S0169-5347(01)02283-2

Eigaard, O.R., Bastardie, F., Hintzen, N.T., Buhl-Mortensen, L.,
Buhl-Mortensen, P., Catarino, R., Dinesen, G.E., Egekvist, J., Fock,
H.O., Geitner, K., Gerritsen, H.D., González, M.M., Jonsson, P.,
Kavadas, S., Laffargue, P., Lundy, M., Gonzalez-Mirelis, G., Nielsen,
J.R., Papadopoulou, N., Posen, P.E., Pulcinella, J., Russo, T., Sala,
A., Silva, C., Smith, C.J., Vanelslander, B., Rijnsdorp, A.D., 2017. The
footprint of bottom trawling in European waters: Distribution,
intensity, and seabed integrity. ICES J. Mar. Sci.
https://doi.org/10.1093/icesjms/fsw194

Garcia, S.M.M., Zerbi,{~ }a., Aliaume, C., Do Chi, T., Lasserre, G.,
2003. The ecosystem approach to fisheries. FAO Fish. Tech. Pap. 443, 71.
https://doi.org/10.1111/j.1467-2979.2010.00358.x

Gladstone-Gallagher, R. V., Pilditch, C.A., Stephenson, F., Thrush,
S.F., 2019. Linking Traits across Ecological Scales Determines
Functional Resilience. Trends Ecol. Evol.
https://doi.org/10.1016/j.tree.2019.07.010

Groenewold, S., Fonds, M., 2000. Effects on benthic scavengers of
discards and damaged benthos produced by the beam-trawl fishery in the
southern North Sea. ICES J. Mar. Sci. 57, 1395--1406.
https://doi.org/10.1006/jmsc.2000.0914

Hiddink, J.G., Jennings, S., Sciberras, M., Bolam, S.G., Cambiè, G.,
McConnaughey, R.A., Mazor, T., Hilborn, R., Collie, J.S., Pitcher, C.R.,
Parma, A.M., Suuronen, P., Kaiser, M.J., Rijnsdorp, A.D., 2019.
Assessing bottom trawling impacts based on the longevity of benthic
invertebrates. J. Appl. Ecol. https://doi.org/10.1111/1365-2664.13278

Hiddink, J.G., Johnson, A.F., Kingham, R., Hinz, H., 2011. Could our
fisheries be more productive? Indirect negative effects of bottom trawl
fisheries on fish condition. J. Appl. Ecol. 48, 1441--1449.
https://doi.org/10.1111/j.1365-2664.2011.02036.x

Hiddink, J.G., Kaiser, M.J., Sciberras, M., McConnaughey, R.A., Mazor,
T., Hilborn, R., Collie, J.S., Pitcher, C.R., Parma, A.M., Suuronen, P.,
Rijnsdorp, A.D., Jennings, S., 2020. Selection of indicators for
assessing and managing the impacts of bottom trawling on seabed
habitats. J. Appl. Ecol. https://doi.org/10.1111/1365-2664.13617

Hinz, H., Prieto, V., Kaiser, M.J., 2009. Trawl disturbance on benthic
communities: Chronic effects and experimental predictions. Ecol. Appl.
19, 761--773. https://doi.org/10.1890/08-0351.1

Jennings, S., Kaiser, M.J., 1998a. The Effects of Fishing on Marine
Ecosystems. Adv. Mar. Biol. 34, 201--352.
https://doi.org/10.1016/S0065-2881(08)60212-6

Jennings, S., Kaiser, M.J., 1998b. The Effects of Fishing on Marine
Ecosystems, in: Advances in Marine Biology. pp. 201--352.
https://doi.org/10.1016/S0065-2881(08)60212-6

Jørgensen, L.L., Planque, B., Thangstad, T.H., Certain, G., 2016.
Vulnerability of megabenthic species to trawling in the Barents Sea.
ICES J. Mar. Sci. https://doi.org/10.1093/icesjms/fsv107

Juan, S. De, Cartes, J.E., Demestre, M., 2007. Effects of commercial
trawling activities in the diet of the flat fish Citharus linguatula (
Osteichthyes\,: Pleuronectiformes ) and the starfish Astropecten
irregularis ( Echinodermata\,: Asteroidea ) 349, 152--169.
https://doi.org/10.1016/j.jembe.2007.05.003

King, J.R., McFarlane, G.A., 2003. Marine fish life history strategies:
Applications to fishery management. Fish. Manag. Ecol.
https://doi.org/10.1046/j.1365-2400.2003.00359.x

Knapp, R.A., Matthews, K.R., Sarnelle, O., 2001. Resistance and
resilience of alpine lake fauna to fish introductions. Ecol. Monogr.
https://doi.org/10.1890/0012-9615(2001)071{[}0401:RAROAL{]}2.0.CO;2

McLaverty, C., Eigaard, O.R., Gislason, H., Bastardie, F., Brooks, M.E.,
Jonsson, P., Lehmann, A., Dinesen, G.E., 2020. Using large benthic
macrofauna to refine and improve ecological indicators of bottom
trawling disturbance. Ecol. Indic.
https://doi.org/10.1016/j.ecolind.2019.105811

Olsgard, F., Schaanning, M.T., Widdicombe, S., Kendall, M.A., Austen,
M.C., 2008. Effects of bottom trawling on ecosystem functioning. J. Exp.
Mar. Bio. Ecol. https://doi.org/10.1016/j.jembe.2008.07.036

Queirós,{~ }a. M., Hiddink, J.G., Kaiser, M.J., Hinz, H., 2006. Effects
of chronic bottom trawling disturbance on benthic biomass, production
and size spectra in different habitats. J. Exp. Mar. Bio. Ecol. 335,
91--103. https://doi.org/10.1016/j.jembe.2006.03.001

Queirós, A.M., Birchenough, S.N.R., Bremner, J., Godbold, J.A., Parker,
R.E., Romero-Ramirez, A., Reiss, H., Solan, M., Somerfield, P.J., Van
Colen, C., Van Hoey, G., Widdicombe, S., 2013. A bioturbation
classification of European marine infaunal invertebrates. Ecol. Evol.
https://doi.org/10.1002/ece3.769

Rice, J., Arvanitidis, C., Borja, A., Frid, C., Hiddink, J.G., Krause,
J., Lorance, P., Ragnarsson, S.Á., Sköld, M., Trabucco, B., Enserink,
L., Norkko, A., 2012. Indicators for sea-floor integrity under the
european marine strategy framework directive. Ecol. Indic.
https://doi.org/10.1016/j.ecolind.2011.03.021

Sciberras, M., Hinz, H., Bennell, J.D., Jenkins, S.R., Hawkins, S.J.,
Kaiser, M.J., 2013. Benthic community response to a scallop dredging
closure within a dynamic seabed habitat. Mar. Ecol. Prog. Ser. 480,
83--98. https://doi.org/10.3354/meps10198

Shin, Y.J., Rochet, M.J., Jennings, S., Field, J.G., Gislason, H., 2005.
Using size-based indicators to evaluate the ecosystem effects of
fishing, in: ICES Journal of Marine Science.
https://doi.org/10.1016/j.icesjms.2005.01.004

Smale, D.A., 2008. Ecological traits of benthic assemblages in shallow
Antarctic waters: Does ice scour disturbance select for small, mobile,
secondary consumers with high dispersal potential? Polar Biol.
https://doi.org/10.1007/s00300-008-0461-9

Thrush, S.F., Hewitt, J.E., Gibbs, M., Lundquist, C., Norkko, A., 2006.
Functional role of large organisms in intertidal communities: Community
effects and ecosystem function. Ecosystems.
https://doi.org/10.1007/s10021-005-0068-8

Tiano, J.C., van der Reijden, K.J., O'Flynn, S., Beauchard, O., van der
Ree, S., van der Wees, J., Ysebaert, T., Soetaert, K., 2020.
Experimental bottom trawling finds resilience in large-bodied infauna
but vulnerability for epifauna and juveniles in the Frisian Front. Mar.
Environ. Res. https://doi.org/10.1016/j.marenvres.2020.104964

Tillin, H.M., Hiddink, J.G., Jennings, S., Kaiser, M.J., 2006. Chronic
bottom trawling alters the functional composition of benthic
invertebrate communities on a sea-basin scale. Mar. Ecol. Prog. Ser.
318, 31--45.

Törnroos, A., Bonsdorff, E., 2012. Developing the multitrait concept for
functional diversity: Lessons from a system rich in functions but poor
in species. Ecol. Appl. https://doi.org/10.1890/11-2042.1

Törnroos, A., Pecuchet, L., Olsson, J., Gårdmark, A., Blomqvist, M.,
Lindegren, M., Bonsdorff, E., 2019. Four decades of functional community
change reveals gradual trends and low interlinkage across trophic groups
in a large marine ecosystem. Glob. Chang. Biol.
https://doi.org/10.1111/gcb.14552

Usseglio-Polatera, P., Bournaud, M., Richoux, P., Tachet, H., 2000.
Biological and ecological traits of benthic freshwater
macroinvertebrates: Relationships and definition of groups with similar
traits. Freshw. Biol. https://doi.org/10.1046/j.1365-2427.2000.00535.x

Van Denderen, P.D., Bolam, S.G., Hiddink, J.G., Jennings, S., Kenny, A.,
Rijnsdorp, A.D., Van Kooten, T., 2015. Similar effects of bottom
trawling and natural disturbance on composition and function of benthic
communities across habitats. Mar. Ecol. Prog. Ser.
https://doi.org/10.3354/meps11550

Zuur, A.F., Ieno, E.N., Walker, N.J., Saveliev, A.A., Smith, G.M., 2009.
Mixed effects models and extensions in ecology with R. Springer Verlag.

\end{document}